\documentclass[paperA4,12pt]{article}
\begin{document}
\title{ Unified description of a class of models for Bose-Einstein
correlations in multiple particle production processes}
\author{K.Zalewski\thanks{Partially supported by the KBN grant
2P03B09322}
\\ M.Smoluchowski Institute of Physics
\\ Jagellonian University, Cracow%\thanks{Address: Reymonta 4, 30-059                                              \\and \\ Institute of Nuclear Physics, Cracow
%Krakow, Poland; zalewski@thrisc.if.uj.edu.pl}
\\ and\\ Institute of Nuclear Physics, Cracow}

\maketitle
\begin{abstract}
Numerous models have been proposed to describe the Bose-Einstein correlations in
multiple particle production process. In the present paper we describe a
generalization, which includes many previous models as special cases and,
therefore, can be useful for work of comparison. We apply the powerful methods of
eigenfunction expansions and generating functionals, which often make the
calculations much shorter than in the original papers.
\end{abstract}

\section{Introduction}

Bose-Einstein correlations in multiple particle production processes at high
energy are now much discussed. The reviews \cite{WIH,HEJ,WEI} contain hundreds of
references and many more can be found in the papers quoted there. There is a
variety of models often based on very different physical pictures of the
production process. In the present paper we stress that many of these models
differ only in their choice of a single particle density matrix, further called
input single particle density matrix. Once this choice has been made, the
calculations become model independent. Following the approach from \cite{BZ1,
BZ2} we describe a general model with a free function, which contains all these
models as special cases. Then we show that the further calculations can be
greatly simplified, if one uses eigenfunction expansions for the input single
particle density matrices and suitable generating functionals to summarize the
information about the distributions, which can be compared with experiment. This
paper is essentially a description from the point of view presented in \cite{BZ1,
BZ2} of a class of models including all the multiparticle symmetrization effects.

\section{Models with factorization}

Many models of Bose-Einstein correlations in multiple particle production
processes can be reduced to the following four steps, which we list here
postponing their discussion to the following sections.

\begin{enumerate}
\item Guess an input single particle density matrix
$\rho_1^{(0)}(\vec{p},\vec{p'})$.
\item Construct the n-particle density matrix for distinguishable
particles
\begin{equation}
\rho_n^0(\vec{p}_1,...,\vec{p}_n;\vec{p'}_1,\ldots,\vec{p'}_n) = \prod_{k=1}^n
\rho_1^{(0)}(\vec{p}_k,\vec{p'}_k).
\end{equation}
\item Symmetrize this density matrix in order to obtain a density matrix for
undistinguishable particles. The diagonal elements, which are enough to calculate
all the momentum distributions, are

\begin{equation}
\rho_n(\vec{p}_1,\ldots,\vec{p}_n) = \frac{1}{n!}\sum_{\sigma,\tau}
\rho_n^{(0)}(\vec{p}_{\sigma 1},\ldots,\vec{p}_{\sigma n};\vec{p}_{\tau
1},\ldots,\vec{p}_{\tau n}),
\end{equation}
where each of the summations over $\sigma$ and $\tau$ extends over all the $n!$
permutations of the set of indices $1,\ldots,n$. The normalization is such that
calculating the trace one integrates over the momentum space without introducing
the factor $\frac{1}{n!}$.
\item Build the diagonal elements of the overall density matrix according to
the formula

\begin{equation}
\rho = \sum_{n=1}^\infty p^{(0)}(n) \rho_n,
\end{equation}
The coefficients are usually chosen Poissonian with some average $\nu$:

\begin{equation}
p^{(0)}(n) = \frac{\nu^n}{n!} e^{-\nu}.
\end{equation}
\end{enumerate}
Further, models built according to this recipe will be called models with
factorization. Let us now discuss the four steps.

\section{Step 1 -- choice of the input single particle density matrix}

The superscript of the function $\rho_1^{(0)}$ indicates that this is not the
observed one-particle density matrix, but an input function necessary for the
construction of the true density matrix. It may be interpreted as the single
particle density matrix for the unphysical case, when the particles are produced
independently and there is no Bose-Einstein symmetrization. The different models
differ in the inspirations used to guess this function. The shortest way is, of
course, to guess directly the input density matrix $\rho_1^{(0)}$, however, a
longer way making use of a model may be easier. Let us quote some examples.

One can guess a source function $S(X,K)$ and calculate the density matrix
$\rho_1^{(0)}$ from the formula

\begin{equation}
\rho_1^{(0)}(\vec{p},\vec{p'}) = \int d^4X e^{-iqX} S(X,K),
\end{equation}
where

\begin{eqnarray}
K& = &\frac{1}{2}(p + p')\\
q& = & p - p'
\end{eqnarray}
are four-vectors with the four-vectors $p,p'$ being on shell particle momenta.
The advantage of this approach is that the source function reflects the
space-time and momentum distribution of the sources of particles. Thus, there are
intuitions what it should look like \cite{WIH,PRA,BIZ}.

Another approach is to assume that the particles originate from a large number
$N$ of independent, incoherent sources. In order to produce a reasonable number
of particles, each of the sources must be weak. A formula for the "total source"
\cite{GKW,HSZ} with a good high $N$ limit is

\begin{equation}
\label{curren} J(\vec{p},N,\zeta, \phi) = \frac{1}{\sqrt{N}}\sum_{k=1}^N
e^{i\phi_k} e^{ipx_k} j_0(\Lambda_k p),
\end{equation}
where $\phi = \{\phi_1,\ldots,\phi_N\}$ is a set of random phases, $\zeta = \{
x_1,\vec{v}_1,\ldots,x_N,\vec{v}_N\}$ is a set of parameters characterizing the
$N$ sources and $\Lambda_k$ are Lorentz transformations with velocities
$\vec{v}_k$, acting on the momentum fourvector $p$. The "total source" is not a
source in the sense of quantum field theory. It is rather a kind of
single-particle wave function with the condition that the components labelled by
different indices $k$ are not allowed to interfere. This is implemented by the
random phases $\phi_k$. An object of this kind can be replaced by a single
particle density matrix

\begin{equation}
\rho_1'(\vec{p},\vec{p'},\zeta) = \frac{1}{N}\sum_{k=1}^N
e^{ix_k(p-p')}j_0(\Lambda_k p) j_0^*(\Lambda_k p').
\end{equation}
There are two ways of proceeding further. One can assume that the density matrix
$\rho_1^0$ is an average of this density matrix over the sets $\zeta_k =
\{x_k,\vec{v}_k\}$ and obtain \cite{HSZ}

\begin{equation}
\rho_1^{(0)}(\vec{p},\vec{p'}) = \int d\zeta_k
\rho(\zeta_k)e^{ix_k(p-p')}j_0(\Lambda_kp)j_0^*(\Lambda_kp').
\end{equation}
In the approach of ref. \cite{HSZ} there are additional factors dependent on the
number of sources $N$, which for $N$ large drop out from the final result. We
absorb the $N$-dependent factor into the density $\rho(\zeta_k)$ so that it does
not appear explicitly in the formulae. In this approach one has to guess the
spectrum of a single source at rest $j_0(p)$ and the distribution of sources
$\rho(\zeta_k)$. The resulting single particle input density matrix does not
depend on the unmeasurable parameters any more.

 Alternatively one can postpone the averaging over the unmeasurable parameters
 as done in ref. \cite{GKW} (further quoted GKW) and choose the pure state input
 single particle density matrix

\begin{equation}\label{}
  \rho_1^{(0)}(\vec{p},\vec{p'}) =
  \frac{1}{\overline{n}(N,\zeta,\phi)}J(\vec{p},N,\zeta,\phi)J^*(\vec{p'},N,\zeta,\phi),
\end{equation}
where

\begin{equation}\label{}
\overline{n}(N,\zeta,\phi) = \int d^3k |J(\vec{k};N,\zeta,\phi)|^2
\end{equation}
is a normalizing constant. It is natural to assume in such models that also the
input multiplicity distribution $p^{(0)}(n)$ depends on the unmeasurable
parameters. Models of this kind do not belong to the class of models discussed in
the present paper, unless simplifying assumptions are made. We will discuss only
the simplest case, when the unmeasurable parameters are fixed or absent, and
consequently no averaging is necessary. This model (cf. \cite{BZ1}) will be
called pure state model, because for it the input single particle density matrix,
which can be written in the form

\begin{equation}\label{pursta}
  \rho_1^{(0)}(\vec{p},\vec{p'}) = \psi(\vec{p})\psi^*(\vec{p'}),
\end{equation}
corresponds to a pure state. There are two reasons to consider this grossly
oversimplified model: it is one of the very few models, where multiparticle
effects can be included analytically and it is a good starting point for the
discussion of the much more important GKW model.

Still another strategy is to guess a set of single particle wave packets
$|\alpha\rangle$ and the distribution of such packets $\rho(\alpha)$ \cite{ZIC,
CSZ}. Then

\begin{equation}
\label{packet} \rho_1^{(0)}(\vec{p},\vec{p'}) = \int d\alpha
\langle\vec{p}|\alpha\rangle \rho(\alpha)\langle\alpha|\vec{p'}\rangle.
\end{equation}
Let us note for further reference that if the states $|\alpha\rangle$ form an
orthonormal set, they are the eigenvectors corresponding to the eigenvalues
$\rho(\alpha)$ of the matrix $\rho_1^{(0)}(\vec{p},\vec{p'})$. Then, usually, the
integration gets replaced by a summation.

Whatever the starting point, very often one finally obtains a Gaussian (cf. e.g.
\cite{PRA,HSZ,ZIC})

\begin{equation}
\label{gaussi} \rho_1^{(0)}(\vec{p},\vec{p'}) = \prod_{i=x,y,z}
\frac{1}{\sqrt{2\pi\Delta_i^2}} \exp\left[ - \frac{K_i^2}{2\Delta_i^2} -
\frac{R_i^2 q_i^2}{2}\right].
\end{equation}
In order not to contradict the Heisenberg uncertainty principle, one must have

\begin{equation}
\Delta_i R_i \geq \frac{1}{2}.
\end{equation}
In the wave packet approach the parameters $R_i$ and $\Delta_i$ are expressed in
terms of other parameters in such a way that this condition is automatically
fulfilled. In other approaches one must impose it as a constraint.

One could include in the exponent a term $is\vec{q}\cdot\vec{K}$ \cite{ZAJ},
where $s$ is a real constant and the factor $i$ is required by the hermiticity of
the density matrix. Since, however, $\vec{K}\cdot\vec{q} = (p_1^2-p_2^2)/2$, this
addition does not affect (in the momentum representation) the density matrix
elements we are interested in \cite{HSZ}. It does affect the deduced size of the
interaction region, but this problem is not discussed in the present paper.

Let us make some remarks about normalization. It is often convenient to use the
normalization

\begin{equation}
\label{norrho} \int d^3p \rho_1^{(0)}(\vec{p},\vec{p}) = 1.
\end{equation}
In other cases an invariant normalization may be preferable

\begin{equation}
\int \frac{d^3p}{E_p} \rho_{inv 1}^{(0)}(\vec{p},\vec{p}) = 1,
\end{equation}
where $E_p = \sqrt{m^2 + \vec{p}^2}$. The relation between the two density
matrices is

\begin{equation}
\rho_{inv 1}^{(0)}(\vec{p},\vec{p'}) = \sqrt{E_pE_{p'}}\rho_1^{(0)}(\vec{p},
\vec{p'})
\end{equation}
and can be used at any time to go from one normalization to the other. For
definiteness we will use the matrix $\rho_1^{(0)}$ normalized according to
(\ref{norrho}).

The calculation of multiparticle effects can be greatly simplified, when for the
input density matrix $\rho^{(0)}_1$ its eigenfunctions $\psi_n(\vec{p})$ and its
eigenvalues $\lambda_n$ are known. These are defined by the equation

\begin{equation}\label{}
  \int d^3p' \rho^{(0)}_1(\vec{p},\vec{p'})\psi_n(\vec{p'}) = \lambda_n
  \psi_n(\vec{p}),
\end{equation}
The eigenfunctions and the eigenvalues are known both for the pure state model
(\ref{pursta}) and for the Gaussian model (\ref{gaussi}). For the pure state
model obviously

\begin{equation}\label{}
  \psi_0(\vec{p}) = \psi(\vec{p})\qquad \mbox{and}\qquad
  \lambda_n = \delta_{n,0}.
\end{equation}
with the remaining eigenfunctions constrained only by the condition that they are
orthogonal to $\psi_0$ and to each other. For the Gaussian model in one dimension
\cite{BZ1, KZ}

\begin{eqnarray}\label{eiggau}
  \psi_n(p) & = &
  \sqrt{\frac{\alpha}{\sqrt{\pi}2^nn!}}\exp\left[-\frac{\alpha^2p^2}{2}\right]H_n(\alpha
  p), \\
  \lambda_n& = &(1-z)z^n,
\end{eqnarray}
where $n=0,1,\ldots$,

\begin{equation}\label{}
  \alpha = \sqrt{\frac{R}{\Delta}};\qquad z = \frac{2R\Delta-1}{2R\Delta+1}.
\end{equation}
In the three-dimensional case the index $n$ becomes the set $\{n_x, n_y, n_y\}$.
The eigenfunctions and the eigenvalues are

\begin{equation}\label{}
\psi_n(\vec{p}) = \psi_{n_x}(p_x)\psi_{n_y}(p_y)\psi_{n_z}(p_z); \qquad \lambda_n
= \lambda_{n_x}\lambda_{n_y}\lambda_{n_z}.
\end{equation}
In terms of its eigenfunctions and eigenvalues the input density matrix is

\begin{equation}\label{expeig}
  \rho^{(0)}_1(\vec{p},\vec{p'}) = \sum_n \lambda_n
  \psi_n(\vec{p})\psi_n^*(\vec{p'}).
\end{equation}
The normalization condition implies that

\begin{equation}\label{}
  \sum_n \lambda_n = 1.
\end{equation}

\section{Step 2 -- independent particles stage}

The density matrix $\rho_n^{(0)}$ is also an auxiliary construct and does not
correspond to an existing physical system. It describes a system of $n$
independent, distinguishable particles. The particles are independent in the
sense that the average of the product of any two single particle operators acting
on different particles is equal to the corresponding product of averages

\begin{equation}
\label{indepe} \langle\hat{O}_1(\vec{p}_1)\hat{O}_2(\vec{p}_2)\rangle_n =
\langle\hat{O}_1(\vec{p}_1)\rangle_n\langle\hat{O}_1(\vec{p}_1)\rangle_n,
\end{equation}
where the averages are defined by the standard formula

\begin{equation}
\langle\hat{O}\rangle_n = Tr\left[\hat{\rho}_n^{(0)}\hat{O}\right].
\end{equation}

One could ask what is the relation between the physical assumption that pions are
emitted independently and the factorization of the weight function in the
definition of the density matrix. E.g. in the wave packet picture does

\begin{eqnarray}
\rho_n(p_1,\ldots,p_n) &=&
\sum_{\alpha_1,\ldots,\alpha_n}|\alpha_1,\ldots,\alpha_n\rangle
\overline{\rho}_n(\alpha_1,\ldots,\alpha_n)\langle\alpha_1,\ldots,\alpha_n|,\\
\overline{\rho}_n(\alpha_1,\ldots,\alpha_n) &=& \prod_{k=1}^n
\overline{\rho}_1(\alpha_k),
\end{eqnarray}
imply that the emission is independent. As easily checked from definition
(\ref{indepe}) the answer is affirmative, if the wave functions $\langle
\vec{p}_1,\ldots\vec{p}_n|\alpha_1,\ldots,\alpha_n\rangle$ are products of single
particle wave functions $\langle \vec{p}_k|\alpha_k\rangle$. This is the case in
step two of the present model. If the $n$-particle wave function corresponds to
correlated particles, the factorizability of the weight function
$\overline{\rho}_n$ does not help. This is the case after the wave functions are
symmetrized, since then the Bose-Einstein correlations appear. It is a matter of
taste, whether these correlations are ascribed to the symmetrization of the
states $|\alpha_1,\ldots,\alpha_n\rangle$, which are part of the density
operator, or to the symmetrization to the external states
$|\vec{p}_1,\ldots,\vec{p}_n\rangle$. The first choice was made e.g. by Zim\'anyi
and Cs\"org\"o \cite{ZIC} and the second e.g. in the classical paper of the
Goldhabers Lee and Pais \cite{GGL}. This is not a physical distinction, however,
because the scalar product defining the $n$ particle wave function does not
depend on whether the first factor, the second factor, or both factors got
symmetrized.

Thus, the model is an independent emission model \cite{BZ1,BZ2,HSZ} in a
well-defined though somewhat formal sense. When $\rho_1^{(0)}$ depends on
unobservable parameters, independence holds for fixed values of these parameters
and would be destroyed, if one averaged over them.

One could define for distinguishable particles, in analogy to step 4, a density
matrix, which involves all the multiplicities

\begin{equation}
\rho^{(0)} = \sum_{n=0}^\infty \frac{\nu^n}{n!} e^{-\nu} \rho_n^{(0)}.
\end{equation}
It is easy to check that also this matrix corresponds to independent particle
production. A different choice of the probabilities $p^{(0)}(n)$ would correspond
to correlated production even at the stage when the particles are considered
distinguishable \cite{HSZ}. Of course, introducing correlations by modifying the
input multiplicity distribution only, without modifying the independent
production assumption for each given multiplicity, is not the most general way of
introducing correlations.

\section{Step 3 -- exclusive momentum distributions}

The diagonal elements of the symmetrized $n$-particle density matrix obtained in
step 3 yield the momentum distribution for $n$ identical particles, when no more
particles of this kind have been produced. There are no constraints on the
production of particles of other kinds. Thus, strictly speaking, this is a
semiinclusive distribution. Since, however, in this paper particles of other
kinds are ignored and could just as well be assumed to be absent, we have called
this distribution exclusive. It is in general not normalized, even when the
single particle density matrix $\rho_1^{(0)}$ is normalized. We will use the
notation \cite{BZ1}: $W_0 = 1$ and for $n>0$

\begin{equation}
W_n = Tr\rho_n = \frac{1}{n!}\sum_{\sigma,\tau} \int \prod_{k=1}^n
\rho_1^{(0)}(\vec{p}_{{\sigma 1} k},\vec{p}_{\tau k})d^3p_k
\end{equation}
Using the matrix $\rho^{(0)}_{inv 1}$ instead of $\rho^{(0)}_1$ one can make this
formula invariant. The same remark applies to our further formulae and we will
not repeat it. For further use it is convenient to define the generating
functional \cite{BZ1,BZ2}: $W_0[u]=1$ and for $n>0$

\begin{equation}
W_n[u] = \int \rho_n(p_1,...,p_n)\prod_{k=1}^n u(\vec{p}_k)d^3p_k.
\end{equation}
For $u(\vec{p}) \equiv 1$ one recovers the parameters $W_n$:  $W_n \equiv
W_n[1]$.

Since each permutation can be decomposed into cycles, the parameters $W_n$ for
$n>0$ can be expressed in terms of the simpler parameters $C_k$ defined for $k>0$
by

\begin{equation}
C_k = Tr[\rho_1^{(0)}]^k = \int \rho_1^{(0)}(\vec{p}_1,\vec{p}_2)
\rho_1^{(0)}(\vec{p}_2,\vec{p}_3) \ldots \rho_1^{(0)}(\vec{p}_k,\vec{p}_1)
\prod_{j=1}^n d^3p_j.
\end{equation}
If the input matrix $\rho_1^{(0)}$ is normalized by (\ref{norrho}), $C_1=1$. The
corresponding functional is

\begin{equation}
C_k[u]  = \int \rho_1^{(0)}(\vec{p}_1,\vec{p}_2)
\rho_1^{(0)}(\vec{p}_2,\vec{p}_3) \ldots \rho_1^{(0)}(\vec{p}_k,\vec{p}_1)
\prod_{j=1}^n u(\vec{p}_j)d^3p_j.
\end{equation}
with $C_k[1] = C_k$.

The functional $W_n[u]$ for $n>0$ can be expressed in terms of the functionals
$C_k[u]$ according to the formula

\begin{equation}
\label{wcycle} W_n[u] = n! \sum_{n_1,\ldots,n_n}\prod_{k=1}^n
\frac{(C_k[u]/k)^{n_k}}{n_k!}.
\end{equation}
The sum is over all the sets of nonnegative integers $\{n_1,\ldots,n_n\}$
satisfying the relation $\sum_{k=1}^n k n_k = n$, or equivalently over all the
decompositions of the set of permutations of $n$ objects into cycles, so that
there are $n_k$ cycles of length $k$. Note that $Tr\rho^{(0)}_n = C_1^n$ is equal
to the term $n_k = n\delta_{k,1}$ of this sum. Thus, all the further terms in
$W_n[u]$ can be interpreted as corrections due to symmetrization.

At this point one can write down the normalized $n$-particle exclusive momentum
distribution

\begin{equation}
P(\vec{p}_1,\ldots,\vec{p}_n) = \frac{1}{W_n}\rho_n(\vec{p}_1,\ldots\vec{p}_n).
\end{equation}
The distribution is here normalized to unity, but a change of this convention
would be trivial.

Using representation (\ref{expeig}) for the input single particle density matrix,
one can write the parameters $C_k$ in the form

\begin{equation}
\label{cieige} C_k = \sum_n \lambda_n^k.
\end{equation}

For the pure state model model (\ref{pursta})

\begin{equation}\label{}
  C_k[u] = \left(\int d^3p |\psi(\vec{p})|^2u(\vec{p}\right)^k; \qquad k = 1,2,\ldots,
\end{equation}
which implies for each $k$: $C_k = 1$ and $W_k[u] = k!C_k[u]$.

For the Gaussian model in three dimensions, after summing three geometrical
progressions,

\begin{equation}\label{}
C_k = \prod_{i=x,y,z}^3\frac{(1-z_i)^k}{1 - z_i^k}.
\end{equation}

Two limits are here of interest \cite{HSZ,ZIC}. When the phase space per particle
is minimal, for $i=x,y,z$: $2\Delta_i R_i \rightarrow 1$ and $z_i \rightarrow 0$.
Consequently, $C_k \rightarrow 1$ for all $k>0$. The state becomes pure as it
should, when due to Einstein's condensation practically all the particle are in
the single particle state corresponding to the eigenvalues $\lambda_{0i}$. When
the phase space is large, for each $i$: $\Delta_i R_i \rightarrow \infty$ and
$z_i \rightarrow 1$. Consequently, $C_k \rightarrow \delta_{k,1}$. In this limit
multiparticle effects become negligible.

Let us note that in order to calculate momentum distributions one uses only the
diagonal elements of the density matrix. Therefore, for this calculation any
density matrix can be replaced by a diagonal matrix  with the same diagonal
elements. For instance, the density operator proposed in refs \cite{GKW, HSZ}:

\begin{equation}\label{gkwdop}
\hat{\rho} = e^{-\overline{n}}\exp\left[i\int dp
J(\vec{p})a^{\dagger}_{\vec{p}}\right]|0\rangle\langle 0|\exp\left[-i\int dp
J^*(\vec{p})a_{\vec{p}}\right],
\end{equation}
where $a_{\vec{p}},a_{\vec{p}}^{\dagger}$ are annihilation and creation operators
for particles with momentum $\vec{p}$, yields a density matrix nondiagonal in
$n$. All that matters for the calculation of momentum distributions, however, are
the diagonal elements and in our notation one finds for each subspace of
$n$-particle states

\begin{equation}\label{}
p^{(0)}(n)\langle \vec{p}_1,\ldots,\vec{p}_n|\hat{\rho}_n|
\vec{p}_1,\ldots,\vec{p}_n\rangle
 = \frac{e^{-\overline{n}}}{n!}\prod_{k=1}^n|J(\vec{p}_k)|^2.
\end{equation}
Thus choosing $p^{(0)}(n) = \exp(-\overline{n})\overline{n}^n/(n!)^2$ one can
replace the density operator (\ref{gkwdop}) by the much simpler density operator
of the pure state case.  For given multiplicity the probability distribution for
the particle momenta is a product of single particle momentum distributions.
Therefore, particle momenta are uncorrelated. In particular, no Bose-Einstein
correlations are seen in the distributions of relative momenta. This corresponds
to the coherent case of the GKW model \cite{GKW}. If, on the other hand, the
unobservable parameters have not been averaged out, the (unnormalized) exclusive
momentum distribution is

\begin{equation}\label{}
  \langle p^{(0)}(n)\rho_n(\vec({p}_1,\ldots,\vec{p}_n)\rangle = \left\langle
  e^{-\overline{n}}\prod_{k=1}^n|J(\vec{p}_k)|^2\right\rangle,
\end{equation}
where $\langle \ldots \rangle$ denotes averaging over the unobservable
parameters. This distribution is not a product of single particle distributions
any more and consequently correlations among the particles are present. It
corresponds to the chaotic case of the GKW model.

\section{Step 4 -- multiplicity distribution}

Let us note first that, since in general $W_n = Tr\rho_n \neq 1$, the
coefficients $p^{(0)}(n)$ are not the probabilities for producing $n$ particles.
These (unnormalized) probabilities are

\begin{equation}
p(n) = p_n^{(0)}W_n.
\end{equation}
Substituting the Poisson formula for $p^{(0)}(n)$ and formula (\ref{wcycle}) with
$u \equiv 1$ for $Tr\rho_n$ one finds

\begin{equation}\label{sumapn}
\sum_n p(n) = e^{-\nu}\exp\left[\sum_{k=1}^\infty \frac{C_k}{k}\nu^k\right].
\end{equation}
For non-Poissonian input multiplicity distributions the corresponding summation
usually must be done numerically \cite{HSZ}. In the following we limit our
discussion to the Poissonian case.  The probability distribution $p(n)$ can be
normalized, if and only if the series in the exponent in (\ref{sumapn})
converges. This is the case if

\begin{equation}
\nu \lim_{k\rightarrow \infty} \sqrt[k]{C_k} = \nu \lambda_0 < 1,
\end{equation}
where $\lambda_0$ is the largest eigenvalue of the input single particle density
matrix. The series diverges when $\nu
> \lambda_0^{-1}$. This happens for the pure state model when

\begin{equation}\label{}
  \nu > 1
\end{equation}
and for the Gaussian model, when

\begin{equation}
\nu > \nu_0 = \prod_{i=x,y,z}(\frac{1}{2} + \Delta_i R_i).
\end{equation}
In the terminology of thermodynamics $\nu = \nu_0$ corresponds to a singularity
of the (grand)partition function, i.e. (usually, see below) to a phase
transition. We show further that this is Einstein's condensation. Note that we
have given here a complete derivation of the formula for $\nu_0$ in the Gaussian
case. Without using the representation of the input density matrix in terms of
its eigenvalues, this calculation would have taken many pages (\cite{ZIC} and
references given there). As Zim\'anyi and Cs\"org\"o put it \cite{ZIC} "these
solutions are not easily obtained" --- and they worked for simplicity with the
spherically symmetric case, where $\Delta_i$ and $R_i$ do not depend on $i$. For
$\nu < \nu_0$ the normalized probability distribution for finding $n$ particles
is

\begin{equation}
P(n) = \frac{\nu^n}{n!}W_n\exp\left[-\sum_{k=1}^\infty \frac{C_k}{k} \nu^k
\right].
\end{equation}
When $\nu$ tends to $\nu_0$ from below, the sum in the exponent of the
normalizing factor tends to infinity and the probability of finding $n < n_0$
particles for any finite $n_0$ tends to zero. This effect disappears, when the
eigenvalue $\lambda_0$ and the corresponding eigenstate $\psi_0$ are removed from
the density matrix (\ref{expeig}). Thus, the surplus of particles condenses in
the state $\psi_0$ -- Einstein's condensation occurs. Since in the pure state
model all the particles are in one pure state anyway, there is no Einstein
condensation in this model for $\nu \rightarrow 1$.

For comparison with experiment one needs the multiplicity distribution, its
moments, cumulants etc, as well as inclusive single particle, two particle etc.
distributions. All this information is conveniently summarized in the generating
functional\footnote{Many of the steps below become obvious, when one translates
the argument into the language of thermodynamics \cite{BZ3}.}

\begin{equation}
\Phi[u] = \sum_{n=0}^\infty p^{(0)}(n)W_n[u] = \exp\left[\sum_{k=1}^\infty
\frac{C_k[u]}{k}\nu^k - \nu\right].
\end{equation}
This formula differs from the corresponding formula in ref. \cite{BZ2} by a
momentum independent factor, which does not affect the logarithmic derivatives of
interest.  For the pure state model

\begin{equation}\label{}
  \Phi[u] = \frac{e^{-\nu}}{1 - \nu\int d^3p|\psi(\vec{p})|^2u(\vec{p})}
\end{equation}

In order to calculate the moments and/or cumulants of the multiplicity
distribution, one chooses the function $u$ constant. Let us denote this constant
by $z$. Then

\begin{eqnarray}
W_n[z] &=& z^nW_n,\\
C_k[z] &=& z^kC_k
\end{eqnarray}
Let us calculate for example

\begin{eqnarray}
\langle n \rangle &=& \left(\frac{\partial\log\Phi[z]}{\partial z}\right)_{z=1} =
\sum_{k=1}^\infty \nu^k C_k,\\
\langle n^2 \rangle - \langle n \rangle^2  &=&
\left(\frac{\partial^2\log\Phi[z]}{\partial z^2}\right)_{z=1}+\langle n \rangle =
\sum_{k=1}^\infty \nu^kk C_k = \nu\frac{d}{d\nu}\langle n \rangle.
\end{eqnarray}
These formulae can be simplified by using the eigenvalue expansion. For instance,
assuming that the geometrical series converge, one finds

\begin{equation}
\label{naveig} \langle n \rangle = \sum_{n=0}^\infty
\frac{\nu\lambda_n}{1-\nu\lambda_n}.
\end{equation}
For given $\rho_1^{(0)}$, i.e. for given eigenvalues $\lambda_k$, the parameter
$\nu$ should be chosen so as to reproduce the observed value of $\langle n
\rangle$. With increasing $\langle n \rangle$ the parameter $\nu$ increases. For
$\langle n \rangle \rightarrow \infty$, $\nu\lambda_0 \rightarrow 1$. Each term
in the sum (\ref{naveig}), except for the first one, tends to a finite limit. Let
us assume that the sum of these limits is non zero and finite. Then, with
$\langle n \rangle$ increasing to large values, only the first term keeps growing
significantly. The $k$-th term in the sum gives the average population of the
$k$-th eigenstate of the density operator. Thus for $\langle n \rangle$ large and
growing, almost all the additional particles due to the increase of $\nu$ land in
one state ($|\psi_0\rangle$) --- Einstein's condensation occurs. The formula for
the dispersion of the multiplicity distribution can be rewritten as

\begin{equation}
\langle n^2 \rangle - \langle n \rangle^2 = \sum_{n=0}^\infty
\frac{\nu\lambda_n}{(1-\nu\lambda_n)^2}
\end{equation}

In the pure state model, where each of the coefficients $C_k$ equals one, the
series for $\langle n\rangle$ converges only if $\nu < 1$ and then

\begin{equation}\label{}
\langle n \rangle = \frac{\nu}{1-\nu}\qquad \mbox{for}\qquad \nu < 1.
\end{equation}
For this model symmetrization from step 3 reduces to a multiplication of the
density matrix by $n!$. The unnormalized probability for producing $n$ particles

\begin{equation}\label{}
  p(n) = e^{-\nu}\nu^n,
\end{equation}
which explains the singularity in $\langle n \rangle$ for $\nu \rightarrow 1$.

As seen from these examples, when the eigenvalues $\lambda _n$ are known, the
numerical evaluation of the moments of the particle multiplicity distribution is
easy.

\section{Step 4 -- inclusive distribution of momenta}

When calculating the inclusive distribution of momenta it is convenient to use
the functions

\begin{equation}
L(\vec{p},\vec{p'}) = \sum_{k=1}^\infty \nu^k \left[\rho_1^{(0)}\right]^k
(\vec{p},\vec{p'}),
\end{equation}
where

\begin{equation}
\left[\rho_1^{(0)}\right]^k(\vec{p},\vec{p'}) = \int d^3p_1\ldots d^3p_n
\rho_1^{(0)}(\vec{p},\vec{p}_1)\rho_1^{(0)}(\vec{p}_1,\vec{p}_2)\ldots
\rho_1^{(0)}(\vec{p}_k,\vec{p'}),
\end{equation}
or equivalently

\begin{equation}
\left[\rho_1^{(0)}\right]^k(\vec{p},\vec{p'}) = \sum_n
\psi_n(\vec{p})\lambda_n^k\psi^*_n(\vec{p'}).
\end{equation}
Using the latter notation

\begin{equation}
L(\vec{p},\vec{p'}) = \sum_n \psi_n(\vec{p})\psi_n^*(\vec{p'})\frac{\nu
\lambda_n}{1-\nu\lambda_n}.
\end{equation}
As seen from this formula

\begin{equation}
L^*(\vec{p},\vec{p'}) = L(\vec{p'},\vec{p}).
\end{equation}
Actually, $L(\vec{p},\vec{p'})$ is the symmetrized single particle inclusive
density matrix (see below).

The various inclusive distributions of momenta can be evaluated as functional
derivatives of the generating functional $\Phi[u]$ at $u=1$. Thus, the single
particle distribution and the two particle correlation function are

\begin{eqnarray}
P_1(\vec{p}) &=& \left(\frac{\delta}{\delta u(\vec{p})}\ln\Phi[u]\right)_{u=1} =
L(\vec{p},\vec{p}),\\
P_2(\vec{p}_1,\vec{p}_2) - P(\vec{p}_1)P(\vec{p}_2) &=& \left( \frac{\delta^2}{
\delta u(\vec{p}_1)\delta u(\vec{p}_2)}\ln\Phi[u]\right)_{u=1} = \nonumber \\ &&
= |L(\vec{p}_1, \vec{p}_2)|^2.
\end{eqnarray}

For the pure state model $L(\vec{p},\vec{p'}) = \rho_1^{(0)}(\vec{p},\vec{p'})$
and there are no symmetrization effects in the inclusive distribution of momenta.
For the Gaussian model, using the formulae from Section 3, one can obtain the
function $L(\vec{p},\vec{p'})$ as a power series in $\nu$. The coefficients of
this series are explicitly known Gaussians.

\section{Conclusions}

Let us summarize our conclusions and add a few comments.

\begin{itemize}

  \item Many models used for the description of Bose-Einstein correlations in high
  energy mutiple-particle-production processes belong to the class of factorizable
  models described in the present paper. As example of a model which does not, let
  us quote the string model developed by Andersson and collaborators \cite{ANH, AND, ANR}.
  This model is based on a very different picture of the particle production process,
  though it has been suggested \cite{BOW} that numerically its predictions might be
  very close to the predictions of a suitably chosen factorizable model. In
  practice the models are often supplemented with corrections for resonance
  production, final state strong interactions, Coulomb interactions etc. We have
  not discussed these problems here.

  \item The input consists of an input multiplicity distribution
  (usually Poissonian) and an input single particle density matrix in the
  momentum representation. There is a great variety of physical pictures used to
  suggests these inputs, but there is no consensus on which is the best.

  \item Once the input has been chosen, it is standard to derive general formulae
  for the inclusive and exclusive distributions of momenta, multiplicity
  distributions and various correlation functions and coefficients. In practice,
  however, it is usually difficult to use these formulae beyond the few particle
  distributions uncorrected for multiple particle effects.

  \item The multiparticle effects can be more easily included, when the
  eigenfunctions and eigenvalues of the input single particle particle density
  matrix are known.  Examples are the two cases, where exact solutions are known,
  i.e. the Gaussian model and the pure state model. In general, Einstein's condensation
  occurs, when the model is not the pure state model and the parameter $\nu$ of the input
  Poissonian multiplicity distribution is equal to the inverse of the largest
  eigenvalue of the input single particle density matrix.

\end{itemize}


\begin{thebibliography} {99}

\bibitem{WIH}U.A. Wiedemann and U. Heinz,{\it Phys. Rep. }{\bf 319}(1999)145.
\bibitem{HEJ}U. Heinz and B. Jacak, {\it Ann. Rev. Nucl. Part. Sci.} {\bf
49}(1999)529.
\bibitem{WEI}R.M. Weiner {\it Phys. Rep.} {\bf 327}(2000)250.
\bibitem{BZ1}A. Bialas and K. Zalewski, {\it Eur. Phys. J. } {\bf C6}(1999)349.
\bibitem{BZ2}A. Bialas and K. Zalewski, {\it Phys. Letters } {\bf
B436}(1998)153.
\bibitem{PRA}S. Pratt, {\it Phys. Letters} {\bf B301}(1993)159.
\bibitem{BIZ}A. Bialas and K. Zalewski, {\it Acta Phys. Pol.} {\bf
B30}(1999)359.
\bibitem{GKW}M. Gyulassy, S.K. Kaufmann and L. Wilson, {\it Phys. Rev.} {\bf
C20}(1979)2267.
\bibitem{HSZ}U. Heinz, P. Scotto and Q.H. Zhang, {\it Ann. Phys.} {\bf
288}(2001)325.
\bibitem{ZIC}J. Zim\'anyi and T. Cs\"org\"o, {\it Heavy Ion Physics} {\bf
9}(1999)241.
\bibitem{CSZ}T. Cs\"org\"o and J. Zim\'anyi, {\it Phys. Rev. Letters} {\bf
80}(1996)301.
\bibitem{ZAJ}W.A. Zajc, in H.H. Gutbrot and J Rafelski eds., NATO
ASI Series {\bf B303}(1993)435. Plenum (New York).
\bibitem{KZ} K. Zalewski, {\it Acta Phys. Pol.} {\bf B33}(2002)1361.
\bibitem{GGL}G. Goldhaber, S. Goldhaber, W. Lee and A. Pais, {\it Phys. Rev.}
{\bf 120}(1960)300.
\bibitem{BZ3}A. Bialas and K. Zalewski, {\it Acta Physica Slovaca } {\bf
49}(1999)145.
\bibitem{ANH}B. Andersson and W. Hofmann, {\it Phys. Letters} {\bf
B169}(1986)364.
\bibitem{AND}B. Andersson, {\it Acta Phys. Pol.} {\bf
B29}(1998)1885.
\bibitem{ANR}B. Andersson and M. Ringn\'er, {\it Nucl. Phys.} {\bf
B513}(1998)627.
\bibitem{BOW}M.G. Bowler, {\it Phys. Letters} {\bf B185}(1987)205.
\end{thebibliography}
\end{document}